\documentclass[12pt,draftclsnofoot, onecolumn]{IEEEtran}

\ifCLASSINFOpdf
\else
\fi

\usepackage{amsmath,graphicx}
\usepackage{subfigure}
\usepackage{lineno}
\usepackage{cite}
\usepackage{multirow}
\usepackage{stfloats}
\usepackage{amsfonts}
\usepackage{amssymb}
\usepackage{booktabs} 
\usepackage{makecell}
\usepackage{diagbox}
\usepackage{slashbox}
\usepackage{color}
\usepackage{bm}
\usepackage{algorithm}
\usepackage{algorithmic}
\begin{document}
	
	\title{ \LARGE Deep Reinforcement Learning Based on Location-Aware Imitation Environment for RIS-Aided mmWave MIMO Systems}
	
	\author{
		\IEEEauthorblockN{
			Wangyang Xu, Jiancheng An, Chongwen Huang, Lu Gan, and Chau Yuen, \IEEEmembership{Fellow, IEEE}
			\vspace{-3em}}
		\thanks{The work of Prof. Huang was supported by the China National Key R\&D Program under Grant 2021YFA1000072, National Natural Science Foundation of China under Grant 62101492, Zhejiang Provincial Natural Science Foundation of China under Grant R22F0110230, Zhejiang University Education Foundation Qizhen Scholar Foundation, and Fundamental Research Funds for the Central Universities under Grant 2021FZZX001-21. The work of Lu Gan was supported by Yibin Science and Technology Program under Grant 2020FW007. \it (Corresponding author: Lu Gan).}
		\thanks{W. Xu, J. An, and L. Gan are with the School of Information and Communication Engineering, University of Electronic Science and Technology of China (UESTC), Chengdu, Sichuan, 611731, China. Lu Gan is also with the Yibin Institute of UESTC, Yibin, Sichuan, 643000, China (E-mail: wangyangxu@std.uestc.edu.cn; jiancheng\_an@163.com; ganlu@uestc.edu.cn).}
		\thanks{C. Huang is with College of Information Science and Electronic Engineering, Zhejiang University, Hangzhou 310027, China, and with International Joint Innovation Center, Zhejiang University, Haining 314400, China, and also with Zhejiang-Singapore Innovation and AI Joint Research Lab and Zhejiang Provincial Key Laboratory of Info. Proc., Commun. \& Netw. (IPCAN), Hangzhou 310027, China. (E-mail:  HYPERLINK "mailto:chongwenhuang@zju.edu.cn" chongwenhuang@zju.edu.cn).}
		\thanks{C. Yuen is with Engineering Product Development (EPD) Pillar, Singapore University of Technology and Design, Singapore 487372, Singapore (E-mail: yuenchau@sutd.edu.sg).}
	}
	% make the title area
	\maketitle
	
	\begin{abstract}
		Reconfigurable intelligent surface (RIS) has recently gained popularity as a promising solution for improving the signal transmission quality of wireless communications with less hardware cost and energy consumption. This letter offers a novel deep reinforcement learning (DRL) algorithm based on a location-aware imitation environment for the joint beamforming design in an RIS-aided  mmWave multiple-input multiple-output system. Specifically, we design a neural network to imitate the transmission environment based on the geometric relationship between the user's location and the mmWave channel. Following this, a novel  DRL-based method is developed that interacts with the imitation environment using the easily available location information. Finally, simulation results demonstrate that the proposed DRL-based algorithm provides more robust performance without excessive interaction overhead compared to the existing DRL-based approaches.
	\end{abstract}
	
	\begin{IEEEkeywords}
		Reconfigurable intelligent surface, deep reinforcement learning, imitation environment.
	\end{IEEEkeywords}
	
	\IEEEpeerreviewmaketitle

	\section{Introduction}	
	Recently, reconfigurable intelligent surface (RIS) is emerged as a promising technology that significantly improves the system throughput, spectrum efficiency, and energy efficiency of wireless networks \cite{2WQQjoint2019}. An RIS is equipped with a large number of hardware-efficient and nearly passive reflecting elements, which does not employ active radio frequency chains hence significantly reducing the energy consumption and hardware. 
	
	Nevertheless, the joint transmit beamforming and reflection coefficients design constitutes a challenge in RIS-aided mmWave multiple-input multiple-output (MIMO) systems. In \cite{2WQQjoint2019, an1, guo2020weighted}, the joint beamforming has been investigated under the consideration of various optimization objectives and phase shift models, where several algorithms were applied such as semidefinite relaxation (SDR) \cite{2WQQjoint2019}, alternating optimization (AO) \cite{an1}, and block coordinate descent (BCD) \cite{guo2020weighted}. 
	
	Additionally, deep learning (DL) techniques have been employed to gain various advantages, such as end-to-end, model-free, and data-driven optimizations \cite{wangyangxu, DRLhuang, DRL3}. Among these DL techniques, the deep reinforcement learning (DRL) enables efficient algorithm designs by observing the rewards from the environment and solving sophisticated optimization problems in the RIS-aided systems \cite{DRLhuang, DRL3}. Unlike supervised learning, DRL does not require any labels and is capable of adapting dynamic environments. Nonetheless, channel's variation remains a hurdle to the robust performance of the DRL. Furthermore, most DRL-based solutions are designed for the multiple-input single-output (MISO) systems and require the channel state information (CSI) as the actor network input, which is challenging to implement for RIS equipped with many passive elements.
	
	In this letter, we study the problem of joint beamforming in RIS-aided  mmWave MIMO wireless communications and propose a DRL algorithm based on the location-aware imitation environment network (IEN). In contrast to most previous works, the proposed algorithm employs the readily-available user's location information rather than the accurate CSI. In addition, to improve DRL's poor robustness facing diverse channels, a deep neural network (DNN) is built to imitate the actual environment for decreasing the excessive overhead imposed by DRL's interaction with the actual environment. Finally, simulation results are provided to verify our proposed algorithm.

	\section{System Model and Problem Formulation}
	\subsection{System Model}
	\begin{figure}[tbp]
		\centering {
			\begin{tabular}{ccc}
				\includegraphics[width=0.8\textwidth]{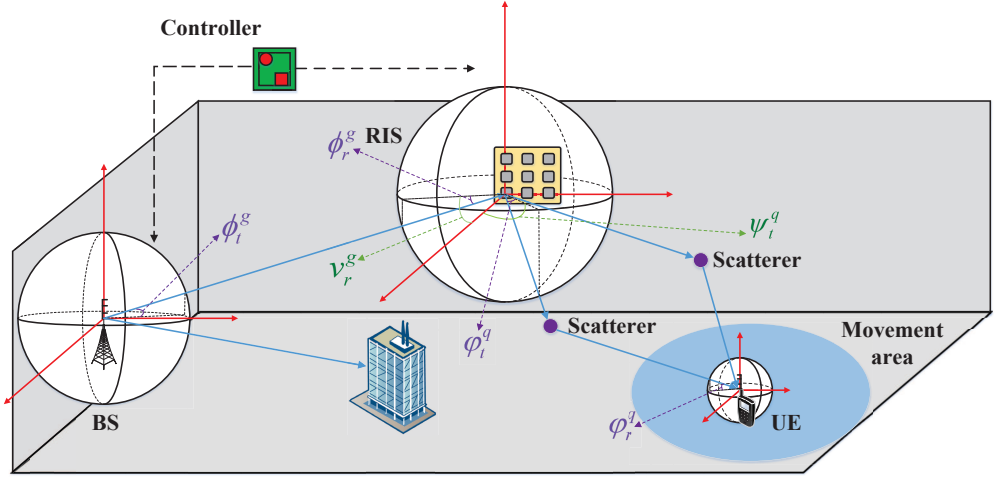}
			\end{tabular}
		}
		\caption{An RIS-aided  mmWave MIMO communication system.}
		\vspace{-1\baselineskip}
		\label{fig1}
	\end{figure}
	
	As illustrated in Fig. \ref{fig1}, we consider a downlink RIS-aided  mmWave MIMO communication system, where an RIS equipped with $N$ passive reflecting elements in a uniform planar array (UPA) is deployed for enhancing the transmission between the base station (BS) with $M$ antennas and the user equipment (UE) with $K$ antennas. The antennas of BS and UE are both arranged in the form of uniform linear array (ULA). 	Each RIS element is capable of rescattering the impinging signals with an individual phase shift, which can be dynamically adjusted by the RIS  controller. Furthermore, we consider the narrowband communications over quasi-static block-fading channels. Let ${\mathbf{G}} \in {\mathbb{C}^{N \times M}}$ and  ${\mathbf{H}_{r}} \in {\mathbb{C}^{K \times N}}$ denote the channel matrices from BS to RIS, and from RIS to UE, respectively. The direct link between the BS and the UE is assumed to be blocked by obstacles.

	Specifically, the mmWave channel is characterized by the classic Saleh-Valenzuela model \cite{wangSVmodel}. Hence, $\bf{G}$ and ${\bf{H}}_{r}$ are generated by 
	\begin{align}   \label{loc1}
		{\bf{G}} = \sqrt {\frac{{MN}}{{{L_G}}}} \sum\limits_{g = 1}^{{L_G}} \sqrt {P{L_g}}  {{\bf{a}}_{P}}\left( {\psi _{G,r}^g,\phi _{G,r}^g} \right){\bf{a}}_L^H\left( {\phi _{G,t}^g} \right),
	\end{align}
	\vspace{-2mm}
	\begin{align}   \label{loc2} 
		{{\bf{H}}_r} = \sqrt {\frac{{NK}}{{{L_D}}}} \sum\limits_{d = 1}^{{L_D}} \sqrt {P{L_d}}  {{\bf{a}}_L}\left( {\phi _{h,r}^d} \right){\bf{a}}_P^H\left( {\psi _{h,t}^d,\phi _{h,t}^d} \right),
	\end{align}
	where $L_G$ ($L_D$) and ${PL _g }$ $({PL _{d}} )$ denote the multi-path number and the complex gain, respectively. ${\psi _{G,r}^g}$ $({\phi _{G,r}^g})$ and ${\phi _{G,t}^g}$$(g  = 1,2, \cdots ,{L_G})$ are the azimuth (elevation) angle of arrival (AoA), and elevation angle of departure (AoD) of the $g$-th path of $\bf{G}$. Meanwhile, ${\phi _{h,r}^d}$ and  ${\psi _{h,t}^d}$ $({\phi _{h,t}^d})$$(d  = 1,2, \cdots ,{L_Q})$ represent the elevation AoA and azimuth (elevation) AoD of ${{{\bf{H}}_r}}$, respectively. ${{\bf{a}}_L}$ and ${\bf{a}}_P$ denote the array response vectors of a half-wavelength spaced ULA and UPA, which are given by
	\begin{align}   \label{loc3} 
		{{\bf{a}}_L}\left( \phi  \right) = \frac{1}{{\sqrt {{N_L}} }}{[1, \cdot  \cdot  \cdot ,{e^{j\pi {n_l}\sin \phi }}, \cdot  \cdot  \cdot ,{e^{j\pi \left( {{N_L} - 1} \right)\sin \phi }}]^T},
	\end{align}
	\vspace{-3mm}
	\begin{align}   \label{loc4} 
		{\bf{a}}_P\left( {\psi  ,\phi } \right) &= \frac{1}{{\sqrt {{N_x}{N_y}} }}\left[ {1, \cdot  \cdot  \cdot ,{e^{j\pi \left( {{n_x}\sin \psi  \cos\phi  + {n_y}\sin \phi } \right)}}, \cdot  \cdot  \cdot ,} \right.\notag\\
		&{\left. {{e^{j\pi \left( {\left( {{N_x} - 1} \right)\sin \psi  \cos\phi  + \left( {{N_y} - 1} \right)\sin \phi } \right)}}} \right]^T,}
	\end{align}
	where $N_L$ is the antenna number of the ULA; $N_x$ and $N_y$ are the number of horizontal and vertical antennas of the UPA; $n_l$, $n_x$ and $n_y$ are the corresponding antenna indices; $\psi $ and $\phi$ are the azimuth and elevation angles, respectively.

	In the downlink transmission phase, the baseband signal ${\mathbf{y}} \in {\mathbb{C}^{K \times 1}}$ received at the UE can be expressed as 
	\begin{align}   \label{loc5}
		{\bf{y}} = {{\bf{H}}_r}{\bf{\Theta G}}{\bf{x}} + {\bf{n}},
	\end{align}
	where ${\bf x}\in {\mathbb{C}^{M \times 1}}$  is the transmitted signal sequence; ${\mathbf{n}} \in {\mathbb{C}^{K \times 1}}$ is the additive white Gaussian noise satisfying ${{\bf{n}}} \sim \mathcal{CN}( {0,{\sigma ^2}{{\bf{I}}_K}} )$; ${\bf{\Theta }} = {\rm{diag}}({\bm{\theta }})$ is the reflection coefficient matrix at the RIS, with ${\bm{\theta }} = {\rm{diag}}({e^{j{\varphi _1}}},{e^{j{\varphi _2}}}, \cdots ,{e^{j{\varphi _N}}})$ and ${\varphi_l}$ is the phase shift of the $n$-th RIS element. ${\rm{diag}}(  \cdot  )$ denotes the diagonal operation. For simplicity, we assume that the phase shifts of each RIS element are continuously adjustable in the interval $\left[ {0,2\pi } \right)$.
	\vspace{-2mm}
	\subsection{Problem Formulation}
	In this letter, we aim to jointly design the transmit signal covariance matrix at the BS and reflection coefficient vector at the RIS for maximizing the achievable rate of RIS-aided  mmWave MIMO systems. Accordingly, the optimization problem can be formulated as
	\begin{align}   \label{loc6}
		\mathop {\max }\limits_{\bf{Q},\bm{\theta} }\;\;  &R = {\log _2}\det \left( {{{\mathbf{I}}_K} + \frac{{{\mathbf{\bar HQ}}{{{\mathbf{\bar H}}}^H}}}{{{\sigma ^2}}}} \right)\notag\\
		s.t.\;\;&{\theta _n} = {{e^{j{\varphi _n}}}}, \varphi _n \in \left[ {0,2\pi } \right),\;\;\;\forall n = 1,2, \cdots ,N,\;\notag\\
		& tr({\bf Q}) \le p, {\bf Q}  \succeq   0,
	\end{align}
	where ${{\bf{\bar H}}}={{\bf{H}}_r}{\bf{\Theta G}}$ is the composite channel, ${\mathbf{Q}} \triangleq \mathbb{E}[{\mathbf{x}}{{\mathbf{x}}^H}]$ denotes the transmit signal covariance matrix, and we consider an average sum power constraint at the transmitter given by $\mathbb{E}[{\left\| {{\mathbf{x}}} \right\|^2}] \le p$, which is equivalent to $tr({\bf Q}) \le p$.
	
	We note that (\ref{loc6}) is a non-convex problem. Although several novel approaches have been proposed for solving this complex problem \cite {zhangshuowen}, the BS generally requires prior knowledge of CSI, even in most data-driven DRL-based algorithms designed for the MISO systems. Furthermore, CSI acquisition remains a tremendous difficulty due to massive passive RIS elements. Besides, the robust performance of these DRL-based algorithms is also poor as UE moves.

	\section{DRL-Based Algorithm with Location-Aware IEN}
\begin{figure}[tbp]
		\centering {
			\begin{tabular}{ccc}
				\includegraphics[width=0.8\textwidth]{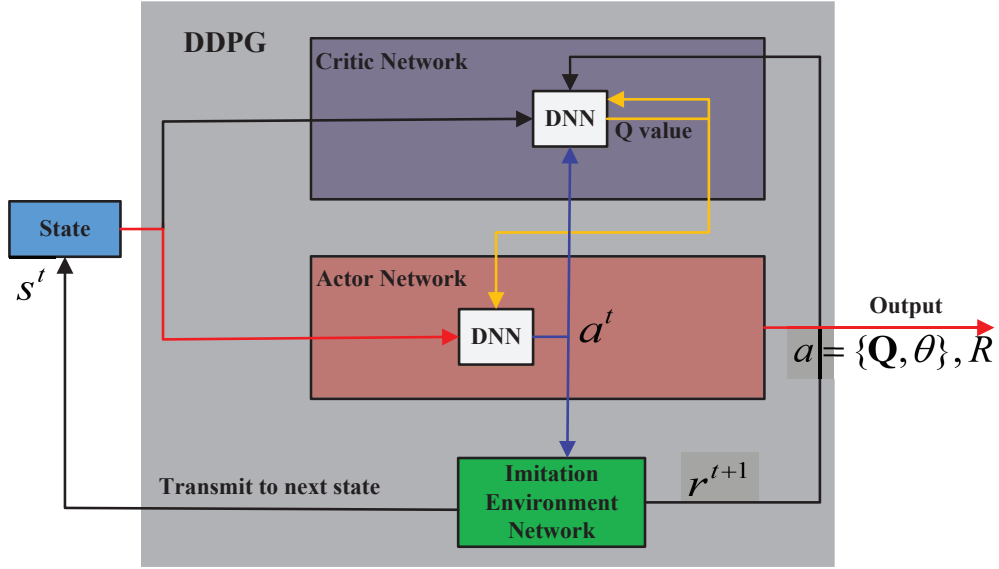}
			\end{tabular}
		}
		\caption{The proposed DRL-based algorithm with IEN.}
		\vspace{-1\baselineskip}
		\label{framework}
	\end{figure}

	In this section, we propose a DRL-based algorithm with location-aware imitation environment for the joint beamforming design of RIS-aided  mmWave MIMO systems. As shown in Fig. \ref{framework}, the proposed DRL-based algorithm mainly uses the off-policy deep deterministic policy gradient (DDPG) network \cite{DDPG}. Besides, an IEN is proposed to replace the actual interaction environment. 
	\vspace{-2mm}
	
	\subsection{The Imitation Environment Network}
	As shown in Fig. \ref{fig1}, BS and RIS are generally placed at fixed positions, leading the slowly time-varying BS-RIS channel $\bf{G}$. By contrast, the RIS-UE channel ${\bf{H}}_r$ varies rapidly owing to the mobility of UE. However, due to the high attenuation in mmWave band, $\bf{G}$ and ${\bf{H}}_r$ rely heavily on the locations of the BS, the RIS, and the UE. As a result, it is possible to recover the transmission channel from the BS to UE by employing only the UE's location information. 
	
	Since the RIS is generally installed at a high place for facilitating LoS propagation with BS/UE, it is reasonable to assume that the RIS serves a single UE with several unknown but fixed scatterers between it and BS/UE. 
	
	First, we build a DNN to imitate the actual transmission environment. Specifically, the IEN consists of two neural networks (NNs), termed the BS-RIS NN and RIS-UE NN, respectively, to provide a general paradigm for different cases of scatterers and NNs input-output mappings. The BS-RIS NN contains three dense layers with 128, 64, and 2$MN$ neurons, respectively. The RIS-UE NN has the same structure as the BS-RIS NN, except that the neurons number of the third dense layer is 2$KN$. The activation functions of the first two dense layers of these two NNs are tanh, while the third one is the linear function. The inputs of these two NNs are the tuples consisting of the 3D coordinates of the devices, $(x_{BS}, y_{BS}, z_{BS}, x_{RIS}, y_{RIS}, z_{RIS})$ and $(x_{RIS}, y_{RIS}, z_{RIS}, x_{UE}, y_{UE}, z_{UE})$. The outputs of the two NNs, ${\rm{vec}}({\mathop{\rm Re}} \{ \bf {\hat G}\} ,{\mathop{\rm Im}}\{ \bf {\hat G}\})$ and ${\rm{vec}}({\mathop{\rm Re}} \{ {\bf {\hat H}}_r\} ,{\mathop{\rm Im}}\{ {\bf {\hat H}}_r\})$, are the vectors of the real and imaginary parts of the predicted $\bf{G}$ and ${\bf{H}}_r$ because the NN can only handle real-valued data. 
	
	Next, we will introduce the details of the training set. We assume that the UE moves in a limited area served by the RIS. Therefore, it is possible to obtain the historical location data of potential UEs by some positioning technologies such as GPS. Specifically, we select $U$ different historic locations of the UE's movement area, and then dynamically and randomly adjust the RIS reflection coefficient vector $F$ times for each location. As a result, we can get a total of $U F$ training inputs in the form of ${\{ lo{c_{BS}},lo{c_{RIS}},loc_{UE}^u,{{\bm{\theta }}_f}\} _{u = 1,2, \cdots ,U;\;f = 1,2, \cdots ,F}}$. Note that $lo{c_{BS}}$, $lo{c_{RIS}}$, and $loc_{UE}^u$ consisting of their 3D coordinates, which will be transformed to the form of the inputs of the BS-RIS NN and BS-RIS NN. Finally, we generate $U F$ corresponding composite channels ${{\bf{\bar H}}}$ as training labels via the traditional channel estimation method \cite{antcom}. We note that there may exist several inaccuracies in the data collection for all locations and composite channels, which will be left for our future research.
	
	During the training phase, we first obtain the outputs, ${\rm{vec}}({\mathop{\rm Re}} \{ \bf {\hat G}\} ,{\mathop{\rm Im}}\{ \bf {\hat G}\})$ and ${\rm{vec}}({\mathop{\rm Re}} \{ {\bf {\hat H}}_r\} ,{\mathop{\rm Im}}\{ {\bf {\hat H}}_r\})$. By reconstructing the complex $\bf {\hat G}$ and ${\bf {\hat H}}_r$, the predicted composite channel is calculated by ${{\bf{\hat H}}} = {{{{\bf{\hat H}}}_r}{\bf{\Theta \hat {G}}}}$. The training of the IEN is based on stochastic gradient descent (SGD), and the loss function in terms of the mean squared error (MSE) are defined as 
	\begin{align}   \label{loc7}
		{\rm{MSE}} = \frac{1}{V}\sum\limits_{v = 1}^V {\left\| {{{{\bf{\hat H}}}_v} - {{{\bf{\bar H}}}_v}} \right\|_F^2},
	\end{align}
	where $V$ is the size of a training batch.
	
	It is noted that the goal of the IEN is to learn the composite ${{\bf{\hat H}}}$ from location information, not the separated $\bf {\hat G}$ and ${\bf {\hat H}}_r$. The detailed training process is shown in Algorithm \ref{algorithm1}. After completing the training process, the predicted ${\bf{\hat H}}$ can thus be used to obtain the predicted achievable rate $\hat R$, denoted as 
	\begin{align}   \label{loc8}
		\hat R={\log _2}\det ( {{{\bf{I}}_K} + \frac{1}{{{\sigma ^2}}}{{\mathbf{\hat HQ}}{{{\mathbf{\hat H}}}^H}}}).
	\end{align}
	\begin{algorithm}[tpb] \vspace{-0mm}
		\caption{The Training of IEN}
		\begin{algorithmic}[1]\label{algorithm1}
			\renewcommand{\algorithmicrequire}{ \textbf{Input:}}
			\REQUIRE ${\{ lo{c_{BS}},lo{c_{RIS}},lo{c_{UE}},{\bm{\theta }}, {{\bf{\bar H}}}\} _{v,v = 1,2, \cdots ,UF}}$.
			\renewcommand{\algorithmicrequire}{ \textbf{Output:}}
			\REQUIRE The trained IEN.       
			\FOR {epoch $i = 1,2, \cdots ,E$} 
			\STATE Construct $I{N_{BR}} = {\rm{(}}{x_{BS}}, {y_{BS}}, {z_{BS}}, {x_{RIS}}, {y_{RIS}}, $${z_{RIS}}{\rm{)}}$ and $I{N_{RU}} = {\rm{(}}{x_{RIS}}, {y_{RIS}}, {z_{RIS}}, {x_{UE}}, {y_{UE}}, {z_{UE}}{\rm{)}}$ by $\{lo{c_{BS}},lo{c_{RIS}},lo{c_{UE}}\}$;
			\FOR {$v = 1,2, \cdots ,V$}
			\STATE Input $I{N_{BR,v}}$ to the BS-RIS NN and output ${{{\bf{\hat G}}}_v}$;
			\STATE Input $I{N_{RU,v}}$ to the RIS-UE NN and output ${{{\bf{\hat H}}}_{r,v}}$;
			\STATE Obtain the predicted composite channel by ${{{\bf{\hat H}}}_v} = {{{\bf{\hat H}}}_{r,v}}{\bf{\Theta }}_v{{{\bf{\hat G}}}_v}$;
			\STATE Calculate MSE by (\ref{loc7}) and update parameters by SGD;
			\ENDFOR
			\ENDFOR
		\end{algorithmic} 
	\end{algorithm}
	\vspace{-2mm}
	\subsection{The DRL-Based Algorithm}
	Inspired by the data-driven DRL approaches \cite{DRLhuang, DRL3}, we proposed a novel DRL-based algorithm with the trained IEN, in which the DDPG network is invoked for solving the complex optimization problem (\ref{loc6}).
	
	Before proceeding, let's introduce some elements that the DDPG requires. 
	\begin{itemize}
		\item \textit{State}: a collection of observations that charaterize the environment. The state ${s^t} \in S$ denotes the observation at the time step $t$, where $S$ is the state space. In this letter, state ${s^t}$ is determined by the RIS reflection coefficient vector ${{\bm{\theta }}^t}$ at time step $t$, the transmit covariance matrix ${\bf{Q}}^t$ at the time step $t$, the achievable rate ${R^t}$ at time step $t$, and the location of the BS, the RIS, and the UE. To accommodate the NN, we convert the complex ${{\bm{\theta }}^t}$ and ${\bf{Q}}^t$ into the vectors containing their real and imaginary parts. Therefore, ${s^t}$ can be represented as $({\rm{vec}}({\mathop{\rm Re}} \{ {\bf {Q}}^t\} ,{\mathop{\rm Im}}\{ {\bf {Q}}^t\}), {\mathop{\rm Re}\nolimits} \{ {{\bm{\theta }}^t}\}, {\mathop{\rm Im}\nolimits}\{ {{\bm{\theta }}^t}\}, {R^t}, lo{c_{BS}}, \\ lo{c_{RIS}}, loc_{UE})$.
		\item \textit{Action}: a set of choices. The agent takes an action step by step during the learning process. Once the agent takes an action ${a^t} \in A$ at time step $t$, the state of the environment will transit from the current state $s^t$ to the next state $s^{t+1}$. As a result, a reward $r^t$ will be fed back to the agent. In this letter, the agent is the RIS controller, and action ${a^t}$ is determined by ${\bf{Q}}^{t+1}$ and ${{\bm{\theta }}^{t+1}}$. We also adopt the real-valued form of ${a^t}$ as $({\rm{vec}}({\mathop{\rm Re}} \{ {\bf {Q}}^{t+1}\} ,{\mathop{\rm Im}}\{ {\bf {Q}}^{t+1}\}), {\mathop{\rm Re}\nolimits} \{ {{\bm{\theta }}^{t+1}}\}, {\mathop{\rm Im}\nolimits}\{ {{\bm{\theta }}^{t+1}}\})$, which needs to be scaled to satisfy the constraints in (\ref{loc6}).
		\item \textit{Q value and reward}: The reward $r^t$ measures immediate return from action ${a^t}$ given state ${s^t}$, whereas the Q value function measures potential future rewards which the agent may get from taking action $a$ at the state $s$. In this letter, the reward $r^t$ is the achievable rate ${R^t}$ at time step $t$.
		\item \textit{Experience}: defined as $({s^t},{a^t},{r^{t + 1}},{s^{t + 1}})$.
	\end{itemize}
	The purpose of DDPG is to maximize the output Q value. As shown in Fig. \ref{framework}, the DDPG contains two basic parts, an actor network and a critic network. The actor network $\mu (s;{\bm{\pi} _\mu })$ takes the state as input and outputs the continuous action, which is in turn input to the critic network together with the state. The critics network $Q(s,a;{\bm{\pi} _Q})$ tries to approach the optimal Q value function. ${\bm{\pi} _\mu }$ and ${\bm{\pi} _Q}$ are the parameters of the actor network and the critic network, respectively. In addition, two target networks copied from the actor network and the critic network, ${\mu^{\prime} }(s;{\pi _{{\mu^{\prime} }}})$ and ${Q^{\prime}}(s,a;{\pi _{{Q^{\prime}}}})$, are created for the better convergence of Q value. These two target networks have the same structure with the original two, but with different parameters. To measure the difference between the critic network's predicted value and the actual target value, a loss function is defined as 
	\begin{align}   \label{loc9}
		Loss({\pi _Q}) = \;\frac{1}{V}\sum\limits_{v = 1}^V {({y_v} - Q(s^t,a^t;{\pi _Q})} {)^2},
	\end{align}
	where ${y_v}$ is the actual target value of the $v$-th sample,  defined as 
	\begin{align}   \label{loc10}
		y = {r^{t + 1}} + \tau \mathop {\max }\limits_{a^{\prime}} Q({s^{t + 1}},a^{\prime};{\pi _{Q^{\prime}}}),
	\end{align}
	where $\tau \in (0,1]$ is the discount rate, $a^{\prime}$ denotes the action output by the target actor network at the time step $t$. 
	
	Therefore, the critic network and the actor network can be updated by the SGD, expressed as 
	\begin{align}   \label{loc11}
		\pi _Q^{t + 1} = \pi _Q^t - {\lambda _Q}{\nabla _{{\pi _Q}}}Loss({\pi _Q}),
	\end{align}
	\vspace{-5mm}
	\begin{align}   \label{loc12}
		\pi _\mu ^{t + 1} = \pi _\mu ^t - {\lambda _\mu }{\nabla _\mu }{Q^{\prime}}({s^t},a;{\pi _{Q^{\prime}}}){\nabla _{{\pi _\mu }}}\mu ({s^t};{\pi _\mu }),
	\end{align}
	where ${\lambda _Q}$ and ${\lambda _\mu }$ are the corresponding learning rates, respectively. Moreover, the updates on the target actor network and the target critic network are given as
	\begin{align}   \label{loc13}
		{\pi _{\mu^{\prime}} } &= {\rho _\mu }{\pi _\mu } + (1 - {\rho _\mu }){\pi _{\mu^{\prime}} },
	\end{align}
	\vspace{-6mm}
	\begin{align}   \label{loc133}
		{\pi _{Q^{\prime}}} &= {\rho _Q}{\pi _Q} + (1 - {\rho _Q}){\pi _{Q^{\prime}}},
	\end{align}
	respectively, where ${\rho _\mu }$ and ${\rho _Q}$ are the corresponding learning rates for updating the target actor network and the target critic network. 
	
	However, the long convergence time of DDPG requires a huge number of transmission time slots and thus limits its application in practical communication systems. Unlike other DRL-based algorithms, the proposed algorithm employs the IEN to replace the interaction between the RIS and the actual environment. The detailed steps of the proposed location-aware DRL-based algorithm are shown in Algorithm \ref{algorithm}.
	\begin{algorithm}[t] \vspace{-0mm}
		\caption{The Location-Aware DRL-based Algorithm}
		
		\begin{algorithmic}[1]\label{algorithm}
			\renewcommand{\algorithmicrequire}{ \textbf{Input:}}
			\REQUIRE $lo{c_{BS}},lo{c_{RIS}},lo{c_{UE}}$.
			\renewcommand{\algorithmicrequire}{\textbf{Output:}}
			\REQUIRE The optimal action $a = \{ {\bf{Q}},{\bm{\theta }}\}$ and the maximum\\ achievable rate $R$ of this algorithm.       
			\FOR {episode $j = 1,2, \cdots ,J$} 
			\STATE Randomly generate the initial $({{\bf{Q}}^0},{{\bm{\theta }}^0})$;
			\STATE Input $\{ lo{c_{BS}},lo{c_{RIS}},loc_{UE},{{\bm{\theta }}^0}\}$ to the trained IEN to obtain ${{{\bf{\hat H}}}^0}$, and then calculate ${{\hat R}^0}$ by (\ref{loc8})  with ${{\bf{Q}}^0}$;
			\STATE Initialize state $s^0=({\rm{vec}}({\mathop{\rm Re}} \{ {\bf {Q}}^0\} ,{\mathop{\rm Im}}\{ {\bf {Q}}^0\}), {\mathop{\rm Re}\nolimits} \{ {{\bm{\theta }}^0}\}, $ \\ ${\mathop{\rm Im}\nolimits}\{ {{\bm{\theta }}^0}\}, {{\hat R}^0}, lo{c_{BS}}, lo{c_{RIS}}, loc_{UE})$;
			\STATE Initialize a random process $\zeta \sim \mathcal{CN} $;
			\FOR {$t = 1,2, \cdots ,T$}
			\STATE Update action ${a^t} = \mu ({s^t};{\pi _\mu }) + \zeta$;
			\STATE Obtain ${{\hat R}^t}$ as step 2 by replacing the corresponding inputs of the IEN, and the next state $s^{t+1}$ will be got accordingly. 
			\STATE Store the experience $({s^t},{a^t},{r^{t + 1}},{s^{t + 1}})$ into the experience replay $\mathcal{B}$;
			\STATE Sample $V$ experiences $({s_v},{a_v},{r_{v+1}},{s_{v+1}})$ from $\mathcal{B}$;
			\STATE Calculate the target Q value according to (\ref{loc10});
			\STATE Update the critic network $Q({s},a;{\pi _Q})$ by (\ref{loc11});
			\STATE Update the actor network $\mu ({s};{\pi _\mu })$ by (\ref{loc12});
			\STATE Update the target actor network and the target critic network by (\ref{loc13}) and (\ref{loc133}).
			\ENDFOR
			\ENDFOR
		\end{algorithmic} 
	\end{algorithm}

	\section{Simulation Results}
	This section provides simulation results to verify the effectiveness of the proposed algorithm. We assume that BS is located at (20, 0, 10) m; RIS is deployed at (0, 30, 20) m; UE is distributed in a  circular movement area with a radius of 5 m and the central location of (10, 50, 0) m. Both $loc_{BS}$ and $loc_{RIS}$ are assumed to be unchanged in the training processes of the IEN and the proposed DRL-based algorithm. Besides, $loc_{UE}$ varies in the former and remains constant in the latter. For the sake of brevity, the BS-RIS channel is considered as the LoS propagation. We assume that there are two scatterers distributed between the RIS and the UE. The locations of the two scatterers are (5, 40, 10) m and (5, 45, 5) m, respectively. The distance-dependent path loss of each link is modeled by ${PL} = {C_0}{d^{ - \alpha }}$, where $C_0=-20$ dB denotes the path loss at the reference distance of 1 m, while $d$ and $\alpha$ denote the transmission distance and the path loss exponent, respectively. The path loss exponents of the BS-RIS and RIS-UE links are set to ${\alpha}_B = 2$ and ${\alpha}_R = 2.8$, respectively. Moreover, we set the transmit power at the BS as $p = 20$ dBm, while the average noise power at the UE is set to ${\sigma ^2} =  - 80$ dBm. Both actor and critic networks have four layers: one input layer, two hidden layers, and one output layer, respectively. The input layer of the actor network has $2(2{M^2} + 2N)+10$ neurons, which is changed to $4(2{M^2} + 2N)+10$ in the critic network. The two hidden layers of both networks contain 500 and 300 neurons, respectively. The output layers of both networks have $2{M^2} + 2N$ and 1 neurons, respectively. Furthermore, we exploit tanh as the activation function, which has a bounded range of outputs facilitating subsequent scaling on RIS phase shifts and the transmit covariance matrix.  The learning rates, ${\lambda _Q}$, ${\lambda _\mu }$, ${\rho _\mu }$, and ${\rho _Q}$ are set as $0.001$. The buff size for experience replay $\mathcal{B}$ and the total episodes $J$ are $10000$ and $1000$, respectively. The batch size $V$ is $16$, while the discount rate $\tau=0.99$.

\begin{figure}[!t]
		\centering
		\subfigure[\label{result1}]{\includegraphics[width=8cm]{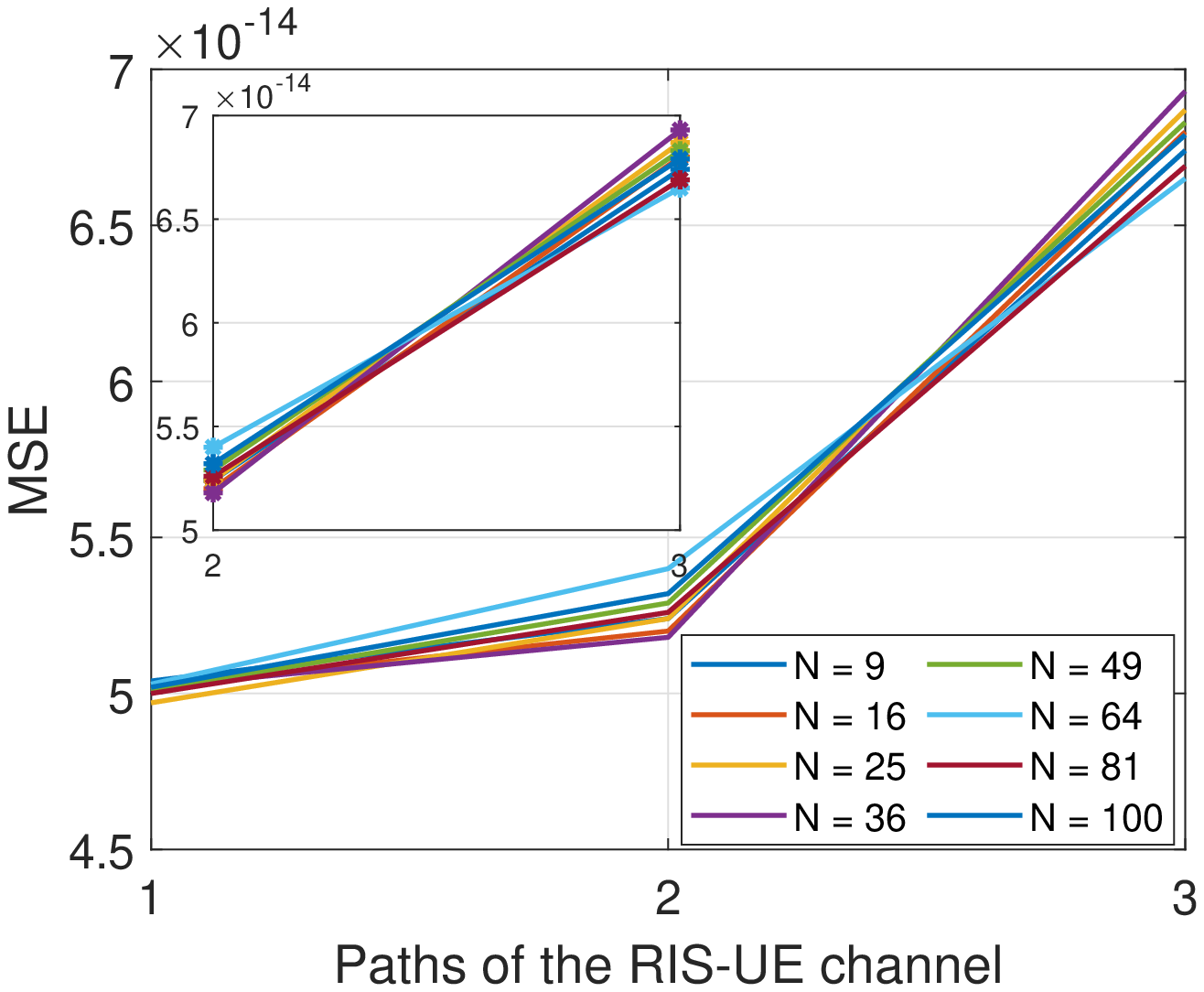}}
		\subfigure[\label{result2}]{\includegraphics[width=8cm]{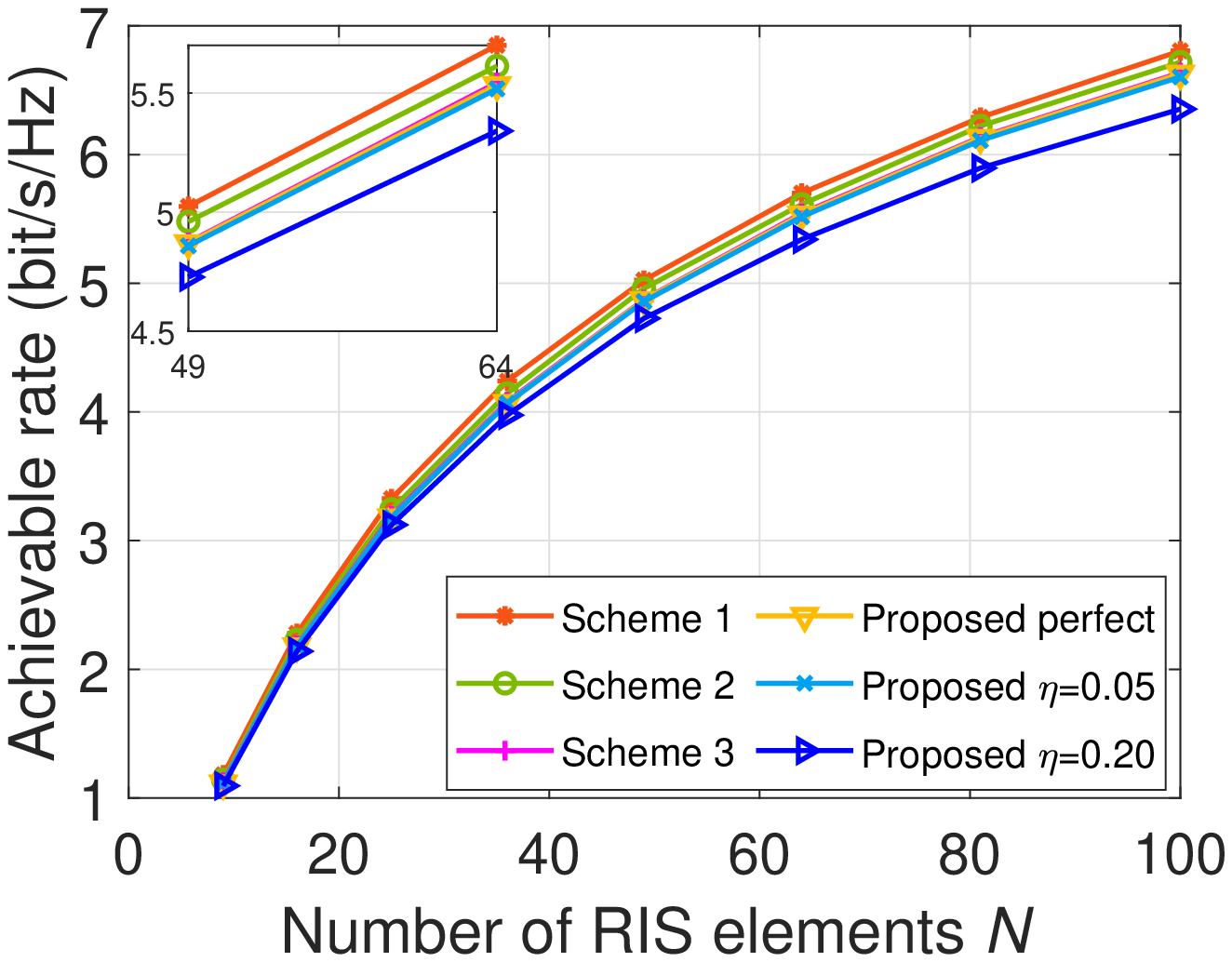}}
		\caption{\color{black}{(a) The MSE of the IEN's output versus the paths number of ${\bf{H}}_r$; (b) The achievable rate versus the number of RIS elements, where we have $M=4$, $K=4$;}}
		\vspace{-0.45cm}
	\end{figure}

	Fig. {\ref{result1}} shows the MSE of the IEN's output versus the paths number of ${\bf{H}}_r$. It can be seen from Fig. {\ref{result1}} that the proposed IEN exhibits similar MSE performance for a given path number with the growing number of RIS elements, which means that the network can adapt to diverse RIS sizes. When the number of RIS elements remains fixed, the fitting performance of the network deteriorates as the number of paths increases, thus necessitating the network to learn more parameters and intrinsic features. Nevertheless, due to the limited number of scatterers in the mmWave band, the proposed network is capable of imitating the actual environment accurately.  
	
	Fig. {\ref{result2}} demonstrates the achievable rate performance versus the number of RIS elements, where we have $M=4$ and $K=4$. For the sake of illustration, we consider three comparison schemes, the AO algorithm (scheme 1) \cite {zhangshuowen}, the CSI-based DRL algorithm (scheme 2) \cite {DRLhuang}, and the location-based DRL algorithm interacting with the actual environment (scheme 3). Observing from Fig. {\ref{result2}}, all of the DRL-based i.e., scheme 2, 3, and the proposed algorithm can achieve comparable achievable rate to the traditional scheme 1. Besides, scheme 3 and the proposed algorithm employing location information perform the same but worse than the scheme 2 utilizing accurate CSI. 
	\begin{figure}[!t]
		\centering
		\subfigure[\label{result3}]{\includegraphics[width=8cm]{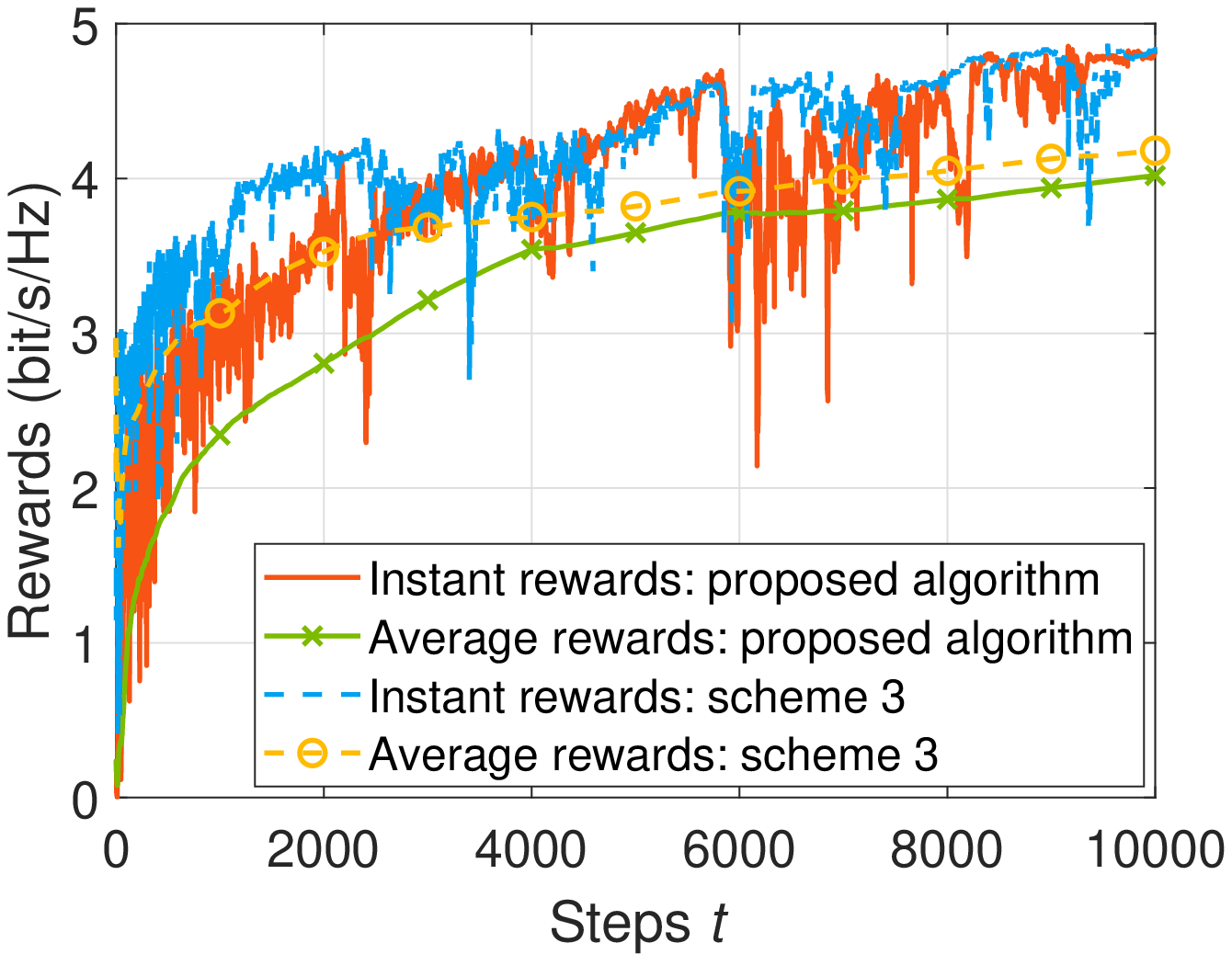}}
		\subfigure[\label{result4}]{\includegraphics[width=8cm]{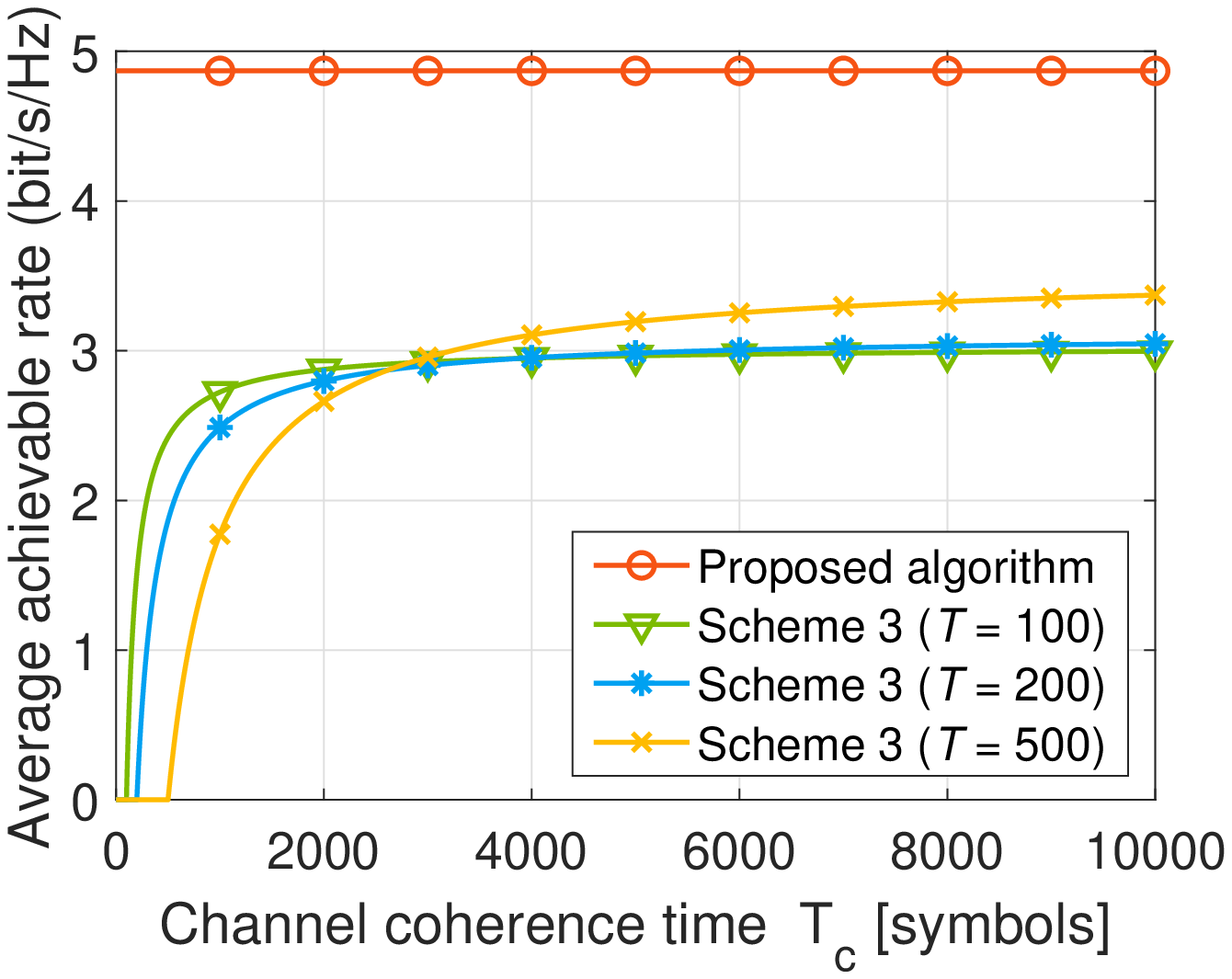}}
		\caption{\color{black}{(a) Rewards versus the steps, where we have $M=4$, $N=49$, $K=4$. (b) The average achievable rate versus the channel coherence time $T_c$, where we have $M=4$, $N=49$, $K=4$.}}
		\vspace{-0.51cm}
	\end{figure}
	The rate performance with the location error of UE is also investigated in Fig. \ref{result2}. We defined the location error as follows, which is caused by the inaccurate positioning method and error during location information transmission.
	\begin{align}   \label{loc14}
		\eta  = {\mathbb{E}{\left[ {\left\| {{\bf{u}} - {\bf{\hat u}}} \right\|} \right]}}/{{\mathbb{E}{\left[ {\left\| {\bf{u}} \right\|} \right]}}},
	\end{align} 
	where ${\bf{u}}=(x_{UE}, y_{UE}, z_{UE})$ is the position vector consisting of the 3D coordinates of the UE, and ${\bf{\hat u}}$ is the corresponding biased position vector. As we can see, the achievable rate of the proposed method decreases as $\eta$ increases. As $N$ increases, the achievable rate gap between the erroneous and perfect cases will gradually widen, which may be caused by the fact that the inaccurate location information can affect more links between the RIS elements and UE as $N$ grows.
	
	In Fig. {\ref{result3}}, we evaluate the instant and average rewards between the proposed algorithm and scheme 3. The average reward ${r_a}(t)$ is defined as 
	\begin{align}   \label{loc15} 
		{r_a}(t) = \frac{{\sum\nolimits_{i = 1}^t {r^t} }}{t},
	\end{align} 
	where ${r^t}$ is the instant reward at the $t$-th time step. We note that these two algorithms converge with a similar trend as the training time increases, and scheme 3 earns a higher reward than the proposed algorithm due to the mismatch of the imitation and actual environment. However, scheme 3 requires many time slots for interacting with the actual environment, whereas the proposed algorithm saves the overhead by interacting with the IEN.
	
	Finally, Fig. {\ref{result4}} compares the average achievable rate of the proposed algorithm and scheme 3 versus the channel coherence time $T_c$. The average achievable rate is defined as 
	\begin{align}   \label{loc16}
		{R_a} = \max (0,\frac{{{T_c} - T}}{{{T_c}}}){R},
	\end{align}
	where $T$ denotes the time slots of signal transmission for interaction, ${R}$ denotes the achievable rate. Fig. {\ref{result4}} shows that the proposed algorithm results in a higher average achievable rate than scheme 3. Since the proposed algorithm interacts with the IEN and saves more time slots for data transmission, specifically, we have $T=0$ for the proposed algorithm, implying that ${R_a}$ remains constant with a given value of training time steps. On the contrary, since scheme 3 requires the signal transmission to interact with the actual environment, we have $T > 0$, which implies that ${R_a}$ of scheme 3 will gradually increase for a growing value of coherence time.

	\vspace{-2mm}
	\section{Conclusion}
	In this letter, a novel DRL-based algorithm was proposed for RIS-aided mmWave MIMO wireless communication systems. In contrast to existing DRL-based algorithms that rely on the perfect CSI, the proposed algorithm employed the readily available UE's location information for circumventing the channel estimation. Furthermore, a network was designed to construct the map between the location information and the composite channel. This network was then employed as an imitation environment of the proposed DRL framework allowing the RIS controller to interact with the UE for the feedback rewards. The IEN decreased the interaction overhead, leading to an average achievable rate enhancement. Simulation results verified the effectiveness and advantages of the proposed algorithm over the existing DRL-based algorithms.

	\ifCLASSOPTIONcaptionsoff
	\newpage
	\fi
	
	\bibliographystyle{IEEEtran}
	\bibliography{Gan_WCL_ref}
\end{document}